\documentclass[epj]{svjour}
\usepackage{graphics}
\usepackage{amsmath}
\usepackage[usenames]{color}
\usepackage{float}
\usepackage[T1]{fontenc}

\begin{document}
\hugehead 
\hyphenation{Gu-ten-berg}
\title{Cross section ratio and angular distributions of the reaction
\boldmath{$\mathrm{p}+\mathrm{d}\rightarrow {^3}{\mathrm{He}}+\eta$}
at $48.8\,\mathrm{MeV}$ and $59.8\,\mathrm{MeV}$
excess energy}
\subtitle{WASA-at-COSY Collaboration}\date{\today}
\author{P.~Adlarson      \inst{1}
 \and W.~Augustyniak     \inst{2} 
 \and{W.~Bardan}         \inst{3}
 \and M.~Bashkanov       \inst{4,5}
 \and F.S.~Bergmann      \inst{6} \fnmsep \thanks{email address: florianbergmann@uni-muenster.de}
 \and M.~Ber{\l}owski    \inst{7}
 \and H.~Bhatt           \inst{8}
 \and M.~B\"uscher       \inst{9,10} \fnmsep \thanks{present address: Peter Gr\"unberg 
  Institut (PGI--6), For\-schungs\-zen\-trum J\"ulich, 52425 J\"ulich, Germany}
 \and H.~Cal\'{e}n       \inst{1} 
 \and I.~Ciepa{\l}       \inst{3}
% ...
}

\institute{Division of Nuclear Physics, Department of Physics and Astronomy, 
  Uppsala University, Box 516, 75120 Uppsala, Sweden 
\and Department of Nuclear Physics, National Centre for Nuclear Research, 
  ul.\ Hoza~69, 00-681, Warsaw, Poland 
\and Institute of Physics, Jagiellonian University, ul.\ Reymonta~4, 30-059 
  Krak\'{o}w, Poland 
\and Physikalisches Institut, Eberhard--Karls--Universit\"at T\"ubingen, Auf 
  der Morgenstelle~14, 72076 T\"ubingen, Germany 
\and Kepler Center f\"ur Astro-- und Teilchenphysik, Physikalisches Institut 
  der Universit\"at T\"ubingen, Auf der Morgenstelle~14, 72076 T\"ubingen, 
  Germany 
\and Institut f\"ur Kernphysik, Westf\"alische Wilhelms--Universit\"at 
  M\"unster, Wilhelm--Klemm--Str.~9, 48149 M\"unster, Germany
\and High Energy Physics Department, National Centre for Nuclear Research, 
  ul.\ Hoza~69, 00-681, Warsaw, Poland
\and Department of Physics, Indian Institute of Technology Bombay, Powai, 
  Mumbai--400076, Maharashtra, India
\and Institut f\"ur Kernphysik, Forschungszentrum J\"ulich, 52425 J\"ulich, 
  Germany
\and J\"ulich Center for Hadron Physics, Forschungszentrum J\"ulich, 52425 
  J\"ulich, Germany
% ...
}

\date{\\}
%
\markboth{P.~Adlarson et al.: Cross section ratio and angular distributions
of the reaction $\mathrm{p}+\mathrm{d}\rightarrow {^3}{\mathrm{He}}+\eta$}
{P.~Adlarson et al.: Cross section ratio and angular distributions of the reaction
$\mathrm{p}+\mathrm{d}\rightarrow {^3}{\mathrm{He}}+\eta$}

\abstract{
We present new data for angular distributions and on the cross section ratio of
the $\mathrm{p}+\mathrm{d}\rightarrow {^3}{\mathrm{He}}+\eta$ reaction at excess
energies of $Q = 48.8\,\mathrm{MeV}$ and $Q = 59.8\,\mathrm{MeV}$. The data have
been obtained at the WASA-at-COSY experiment (Forschungszentrum J\"ulich) using
a proton beam and a deuterium pellet target. While the shape of
obtained angular distributions show only a slow variation with the energy, the
new results indicate a distinct and unexpected total cross section fluctuation
between $Q = 20\,\mathrm{MeV}$ and $Q = 60\,\mathrm{MeV}$, which might indicate
the variation of the production mechanism within this energy interval.
} 
\maketitle
\section{Introduction}
\label{intro} While the very near-threshold region of the reaction \linebreak
$\mathrm{p}+\mathrm{d}\rightarrow {^3}{\mathrm{He}}+\eta$ is well
covered by a broad data set
\cite{Mers2007,Smyrski2007,Berger1988,Mayer1996,Adam2007}, at higher excess
energies only a limited amount of data is available. In more detail,
the total cross section values from ANKE and WASA/PROMICE
\cite{Raus2009,Bilger2002,Bilger2004} expose a plateau in the excitation
function at excess energies between $40\,\mathrm{MeV}$ and
$120\,\mathrm{MeV}$ with the exception of one single data point of
the GEM experiment at $48.8\,\mathrm{MeV}$
\cite{Betigeri2000}. Although still consistent with this cross
section plateau shown by the neighboring data when considering the
statistical and systematic uncertainties, this $48.8\,\mathrm{MeV}$
data point might be a hint for an increase of the cross section
at this energy. Moreover, calculations available in the literature
\cite{Kelkar2013}, based on either a one-step or two-step model, fail to explain
in parallel the forward peaked angular distributions and the total cross section
of this reaction channel. In order to better understand the underlying
production processes and the strong final-state interaction it was proposed to
perform new calculations based on a boson exchange model for which new
high-quality data at intermediate excess energies are required \cite{Kelkar2013}.

Therefore, angular distributions have been
obtained using the WASA-at-COSY installation for the
$\mathrm{p}+\mathrm{d}\rightarrow {^3}{\mathrm{He}}+\eta$ reaction
at excess energies of $Q = 48.8\,\mathrm{MeV}$ and $Q =
59.8\,\mathrm{MeV}$. Furthermore, using the
$\mathrm{p}+\mathrm{d}\rightarrow {^3}{\mathrm{He}}+\pi^0$ reaction the cross
section ratio for the $\eta$ meson production has been determined.

\section{Experiment and Data Analysis}
\label{sec:1}
\begin{figure}
\centering
\resizebox{0.45\textwidth}{!}{
  \includegraphics{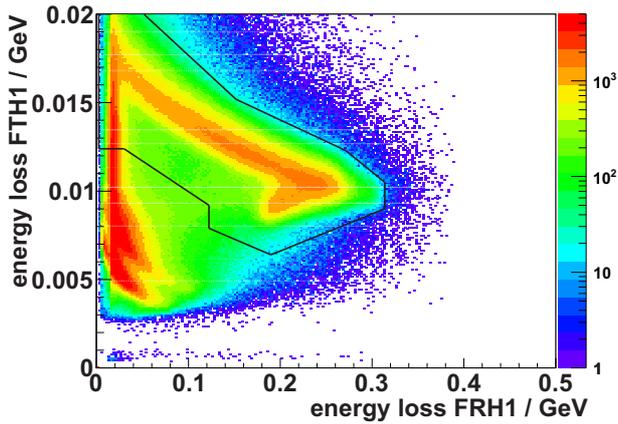}
}
\caption{Energy loss in the first layer of the Forward Trigger Hodoscope (FTH1)
against the energy loss in the first layer of the Forward Range Hodoscope (FRH1)
for the $Q = 59.8\,\mathrm{MeV}$ raw data.}
\label{fig:dEE}
\end{figure}

The experiment was conducted at the COSY storage ring of the 
Forschungszentrum J\"ulich, using the WASA-at-\linebreak
COSY experimental setup \cite{Adam2004}. Protons with beam energies of
$980\,\mathrm{MeV}$ ($Q = 48.8\,\mathrm{MeV}$) and $1000\,\mathrm{MeV}$
($Q = 59.8\,\mathrm{MeV}$),  respectively, were scattered on a deuterium pellet
target \cite{Berg2008} and the produced ${^3}{\mathrm{He}}$ nuclei were stopped
and detected in the Forward Detector. The selection of these ${^3}{\mathrm{He}}$
nuclei is presented in fig.~\ref{fig:dEE}, showing the energy loss in the first
layer of the Forward Trigger Hodoscope (FTH1) against the energy loss in the
first layer of the Forward Range Hodoscope (FRH1). The black solid line presents
the ${^3}{\mathrm{He}}$ selection cut to suppress other ejectiles such as
protons, deuterons and pions \cite{Annika2010}. By this cut much less than one percent of
${^3}{\mathrm{He}}$ nuclei are rejected and, in addition, this factor cancels
out in the following determination of the total cross section ratio.
As expected the energy loss in the thin FTH1 detector decreases with increasing
energy detected in the first layer of the FRH. The kink at a FRH1 energy of
$\approx$ 0.25 GeV is due to the fact that ${^3}{\mathrm{He}}$ nuclei with
higher energies are no longer stopped in the first FRH layer but enter the
second one. However, for the reaction of interest only events stopped in the
FRH1 layer are of relevance. 

\begin{figure}
\centering
\resizebox{0.45\textwidth}{!}{
  \includegraphics{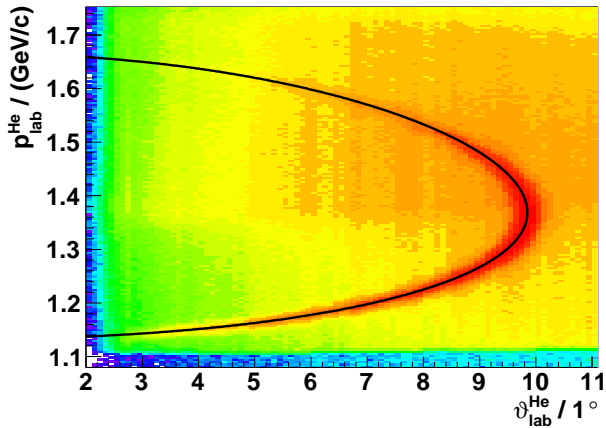}
}
\caption{${^3}{\mathrm{He}}$ laboratory momenta against the corresponding
laboratory angle for the $Q = 59.8\,\mathrm{MeV}$ data compared to the theoretical
expectation (black line).}
\label{fig:momtheta}
\end{figure}

The four-momentum vector of the ${^3}{\mathrm{He}}$ ions is reconstructed by their 
total kinetic energy measured in the forward detector as well as their 
azimuthal and polar scattering angles in the laboratory system. The 
$\mathrm{p}+\mathrm{d}\rightarrow {^3}{\mathrm{He}}+\eta$
reaction is identified by the missing-mass method. The number of
${^3}{\mathrm{He}}$-$\eta$ events was extracted for individual polar angles to
determine angular distributions. Since, for simple kinematic
reasons, for this two-body reaction the particle momenta are directly correlated
with the polar angle in the laboratory system, the energy calibration of the
forward detector could be performed with high accuracy. In detail, the polar angle
$\vartheta_{\mathrm{LAB}}$ is reconstructed with an uncertainty of
$\Delta \vartheta_{\mathrm{LAB}} \approx \pm 0.2\,^{\circ}$, corresponding to,
\emph{e.g.}, a momentum resolution of $\Delta p_{\mathrm{LAB}} \approx 0.5\,\%$ at a
laboratory angle of $\vartheta_{\mathrm{LAB}} = 7\,^{\circ}$. The laboratory
momenta reconstructed from the measured scattering angle have been used for a
careful calibration of the forward detector. Figure \ref{fig:momtheta} shows a
corresponding scatter plot for identified ${^3}{\mathrm{He}}$ nuclei at
$Q = 59.8\,\mathrm{MeV}$. The clear accumulation of events originates from the
$\mathrm{p}+\mathrm{d}\rightarrow {^3}{\mathrm{He}}+\eta$ reaction for which the
theoretical expectation assuming infinite resolution is presented by the solid
line.

\begin{figure}
\centering
\resizebox{0.45\textwidth}{!}{
    \includegraphics{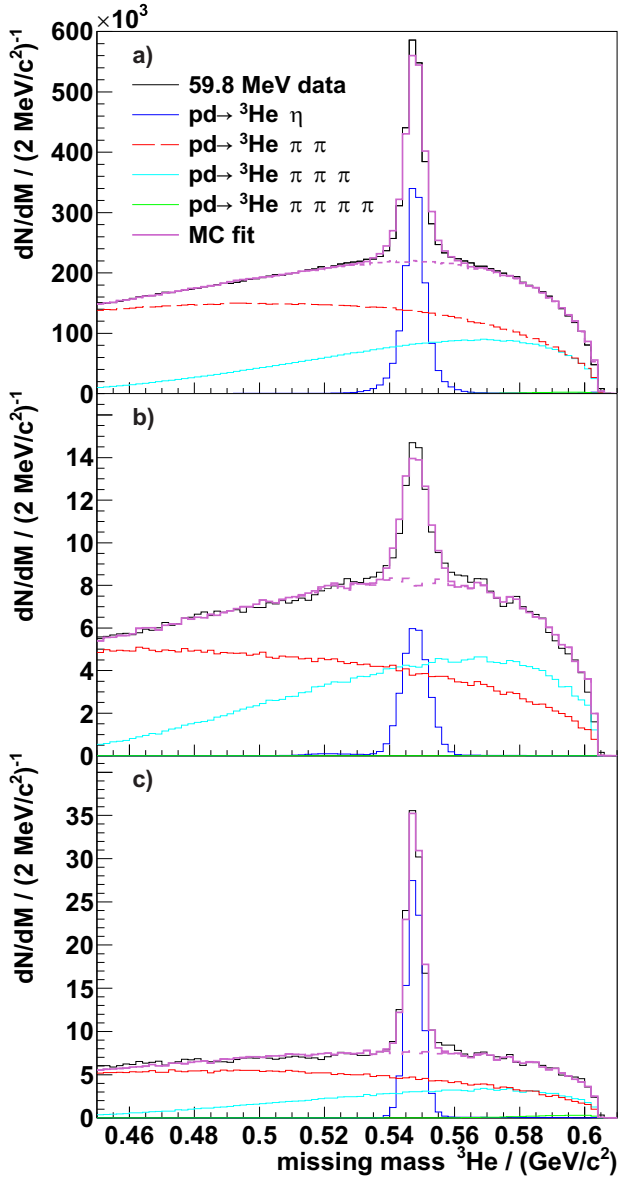}
} \caption{(a) ${^3}{\mathrm{He}}$ missing mass for the $59.8\,\mathrm{MeV}$ data
for the full angular range with a Monte Carlo fit.
(b) ${^3}{\mathrm{He}}$ missing mass for the $59.8\,\mathrm{MeV}$ data
for $-0.44 \leq \cos{\vartheta_{\mathrm{CMS}}^{\eta}} < -0.36$ with a
Monte Carlo fit.
(c) ${^3}{\mathrm{He}}$ missing mass for the $59.8\,\mathrm{MeV}$ data
for $0.36 \leq \cos{\vartheta_{\mathrm{CMS}}^{\eta}} < 0.44$ with a
Monte Carlo fit.
Details of the fits are described in the text.}
\label{fig:missmass}
\end{figure}

For both excess energies the full center-of-mass angular range is divided 
into 25 equally spaced $\cos{\vartheta_{\mathrm{CMS}}^\eta}$ bins for which
the ${^3}{\mathrm{He}}$ missing-mass distribution was determined. The resulting
spectra were then fitted by a Monte Carlo cocktail in the missing mass range
$0.45\,\mathrm{GeV}/{c}^2$--$0.60\,\mathrm{GeV}/{c}^2$, 
considering both the $\eta$ meson production and all energetically allowed 
multi-pion production channels. Figure \ref{fig:missmass} (a) shows the
missing-mass spectrum for the full angular range of the $59.8\,\mathrm{MeV}$ data 
set, while figs.~\ref{fig:missmass} (b) and (c)
show the missing-mass spectra for
$-0.44 \leq \cos{\vartheta_{\mathrm{CMS}}^{\eta}} < -0.36$ and
$0.36 \leq \cos{\vartheta_{\mathrm{CMS}}^{\eta}} < 0.44$, respectively. For the
final analysis this fit was performed for each individual angular 
bin. Uniform phase space distributions were assumed for modeling the missing-mass
distributions and the magnitudes of each multi-pion reaction contribution
were treated as free parameters for each angular bin.
Whereas the four-pion production was found to be of minor relevance,
the two- and three-pion production channels contribute dominantly to the
background description. Although the ABC effect \cite{Bash2006} is known to
strongly influence the shape of the two-pion production background, a very good
background description could be achieved.
Nevertheless the presence of the ABC effect prevents from the direct extraction
of the relative $N_{2\pi} / N_{3\pi}$ ratios.
It should be noted that for angular bins which correspond to the very forward
and backward region the background was fitted by polynomial fits which resulted
in a slightly better description. For better visualization the background
channels with the same number of pions but different charges are merged in this
figure, although every channel was considered separately for the fit. Figures
\ref{fig:missmass}(a), (b), and (c) show that by this method the (multi-pionic)
background can be described well in the vicinity of the $\eta$ mass peak.

\begin{figure}
\centering
\resizebox{0.45\textwidth}{!}{
    \includegraphics{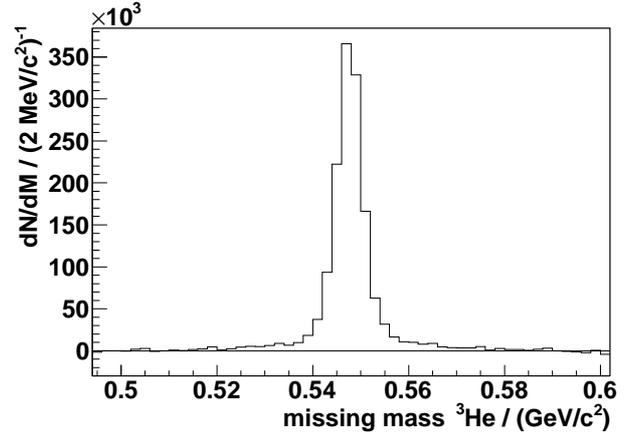}
} \caption{${^3}{\mathrm{He}}$ missing-mass peak for the $59.8\,\mathrm{MeV}$
data for the full angular range after background subtraction.}
\label{fig:mmpeak}
\end{figure}

Figure \ref{fig:mmpeak} shows the total missing-mass spectrum with a 
resolution of $\Delta mm = 3\,\mathrm{MeV}/{\mathrm{c}}^2$ (RMS) after
background subtraction.
The number of $\mathrm{p}+\mathrm{d}\rightarrow {^3}{\mathrm{He}}+\eta$ events
is extracted by considering a $\pm 3 \sigma$ interval. In total $1.3\times10^6$
($1.3\times10^5$) events were selected for the measurement at
$59.8(48.8)\,\mathrm{MeV}$.

\begin{figure}
\centering
\resizebox{0.45\textwidth}{!}{
  \includegraphics{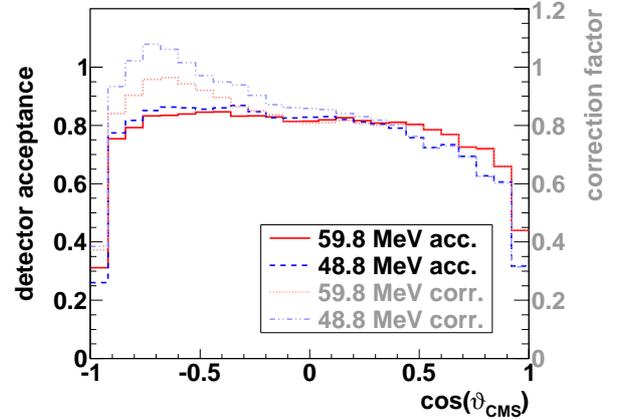}
}
\caption{Monte Carlo simulations of the detector acceptance for the reaction
$\mathrm{p}+\mathrm{d}\rightarrow {^3}{\mathrm{He}}+\eta$
at $48.8\,\mathrm{MeV}$ (dashed line) and $59.8\,\mathrm{MeV}$
(solid line) excess energy and the correction factors
at $48.8\,\mathrm{MeV}$ (dashed dotted line) and $59.8\,\mathrm{MeV}$
(dotted line) excess energy. The statistical uncertainties are in the
order of the line width.
}
\label{fig:etaacc}
\end{figure}

The acceptance of the rotational symmetric forward detector 
for ${^3}{\mathrm{He}}$ nuclei of the $\mathrm{p}+\mathrm{d}\rightarrow
{^3}{\mathrm{He}}+\eta$ reaction was determined by Monte Carlo simulations as
function of the scattering angle (fig.~\ref{fig:etaacc}). The acceptance varies
smoothly in the range of $\epsilon = 60\%$--$86 \%$ with the polar angle. Only 
in the very forward and backward direction a drop of the acceptance is caused
by ${^3}{\mathrm{He}}$ nuclei escaping through the hole for the beam pipe
in the detector. Those two bins have been excluded from the following analysis.

To extract angular distributions the measured distributions
have to be corrected for the geometrical acceptance of the detection system,
for the track reconstruction efficiency as well as for effects caused by the
finite momentum resolution. The first two effects are represented for both
energies by the solid and dashed lines in fig.~\ref{fig:etaacc}. An inclusion of
the latter effect results in the dotted and dashed dotted lines. The presented
correction factors have been determined by an iterative procedure using the
extracted angular distributions as input for the Monte Carlo simulations until
a conversion was reached.

\section{Data normalization}
\label{sec:2}
A relative normalization of the measurements at $48.8\,\mathrm{MeV}$ and
$59.8\,\mathrm{MeV}$ excess energy is obtained from the simultaneous analysis of
the $\mathrm{p}+\mathrm{d}\rightarrow {^3}{\mathrm{He}}+\pi^0$ reaction. The
acceptance and cross section corrected ratio of the respective $\pi^0$ yields
corresponds to the ratio of the integrated luminosities. For both energies
the acceptances were found to be identical within their uncertainties. As for the $\eta$
case, the $\mathrm{p}+\mathrm{d}\rightarrow {^3}{\mathrm{He}}+\pi^0$ reaction is
identified from the missing mass with respect to the ${^3}{\mathrm{He}}$
detected in the forward detector. Due to the higher ${^3}{\mathrm{He}}$ momenta
in the $\pi^0$ production a larger background from protons and deuterons
misidentified as ${^3}{\mathrm{He}}$ is visible. In order to reduce the
background, exactly two photons registered in the central detector from the
$\pi^0 \rightarrow \gamma\gamma$ decay are required. In the analysis
presented here only the polar angle range $-0.92 \leq
\cos{\vartheta_{\mathrm{CMS}}^{\pi^0}} < -0.68$ is used and divided into
three equally sized bins. It should be noted that even when requiring the
$\gamma \gamma$ decay signal the acceptance is above $50\,\%$ for the
chosen angular bins.
\begin{figure}
\centering
\resizebox{0.45\textwidth}{!}{
  \includegraphics{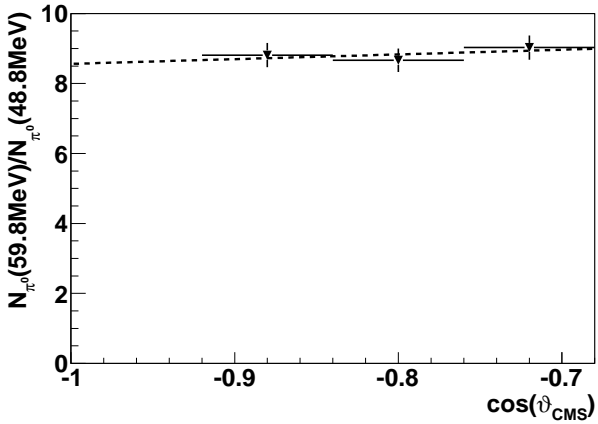}
 } \caption{Relative $\pi^0$ yield for the measurements at $59.8\,\mathrm{MeV}$
and $48.8\,\mathrm{MeV}$ excess energy. The data points refer to the central
values of the chosen angular bins and are shown with statistical uncertainties.
The dashed line shows the linear fit.}
\label{fig:factor}
\end{figure}

\begin{figure}
\centering
\resizebox{0.45\textwidth}{!}{
  \includegraphics{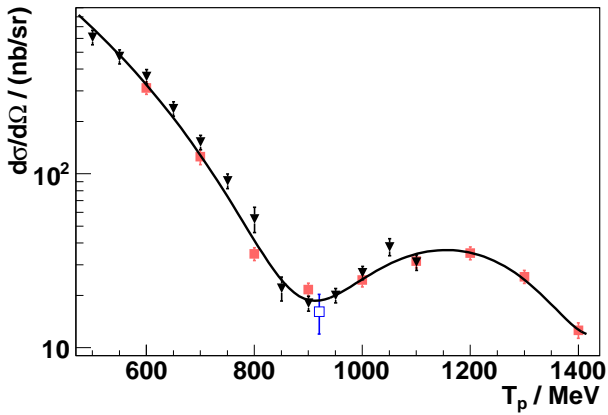}
 } \caption{Differential cross sections for
$\cos{\vartheta_{\mathrm{CMS}}^{\pi^0}} = -1$ as a function of the proton beam
energy $T_{\mathrm{p}}$. The data are taken from ref.~\cite{Berthet1985} for the
reactions $\mathrm{p}+\mathrm{d}\rightarrow {^3}{\mathrm{He}}+\pi^0$ (open
square) and $\mathrm{p}+\mathrm{d}\rightarrow {^3}{\mathrm{H}}+\pi^+$ (filled
squares, scaled by an isospin factor of 0.5) as well as from
ref.~\cite{Kerboul1986} for the
$\mathrm{d}+\mathrm{p}\rightarrow {^3}{\mathrm{He}}+\pi^0$ reaction (triangles)
and fitted by a fifth-order polynomial. The shown error bars include
statistical and systematic uncertainties.}
\label{fig:crosspi0}
\end{figure}

Figure \ref{fig:factor} shows the ratio of the $\pi^0$ yields as function of the
polar angle including statistical uncertainties. In order to extract the ratio
of the $\pi^0$ yields at $\cos{\vartheta_{\mathrm{CMS}}^{\pi^0}} = -1$ the data
was fitted by a linear function. A value of
\begin{eqnarray}
\label{eq:npiratio}
\frac{N_{\pi^0}(T_{\mathrm{p}} = 1000\,\mathrm{MeV})}
{N_{\pi^0}(T_{\mathrm{p}} = 980\,\mathrm{MeV})} = 8.6 \pm 0.6
\end{eqnarray}
was determined with an uncertainty which is dominated by the lower statistics of
the $Q=48.8\,\mathrm{MeV}$ data. For a relative normalization of the $\eta$
cross sections via the
$\mathrm{p}+\mathrm{d}\rightarrow {^3}{\mathrm{He}}+\pi^0$ reaction the ratio
$\sigma_{\pi^0}(T_{\mathrm{p}} = 980\,\mathrm{MeV})/\sigma_{\pi^0}(T_{\mathrm{p}} = 1000\,\mathrm{MeV})$
is needed in addition. Differential cross sections for
$\cos{\vartheta_{\mathrm{CMS}}^{\pi^0}} = -1$ of the reactions
$\mathrm{p}+\mathrm{d}\rightarrow {^3}{\mathrm{He}}+\pi^0$ and
$\mathrm{p}+\mathrm{d}\rightarrow {^3}{\mathrm{H}}+\pi^+$, scaled by an isospin
factor of $0.5$, from ref.~\cite{Berthet1985} as well as differential cross
sections of the $\mathrm{d}+\mathrm{p}\rightarrow {^3}{\mathrm{He}}+\pi^0$
reaction from ref.~\cite{Kerboul1986} have been fitted by a fifth-order
polynomial to extract the required values. Based on this fit shown in
fig.~\ref{fig:crosspi0},
\begin{eqnarray}
\label{eq:sigmapiratio}
\frac{\sigma_{\pi^0}(T_{\mathrm{p}} = 980\,\mathrm{MeV})}
{\sigma_{\pi^0}(T_{\mathrm{p}} = 1000\,\mathrm{MeV})} = 0.914 \pm 0.009
\end{eqnarray}
has been obtained. The quoted uncertainty considers both systematic and
statistical uncertainties. Using eqs.~(\ref{eq:npiratio}) and
(\ref{eq:sigmapiratio}), the cross section ratio for the $\eta$ meson
production is then given by:
\begin{eqnarray}
\label{eq:ratio}
\frac{\sigma_{\eta}(48.8\,\mathrm{MeV})}{\sigma_{\eta}(59.8\,\mathrm{MeV})}
&=& (0.914 \pm 0.009) \cdot (8.6 \pm 0.6) \nonumber \\
&\cdot& \frac{N_{\eta}(48.8\,\mathrm{MeV})}{N_{\eta}(59.8\,\mathrm{MeV})} \ .
\end{eqnarray}
Here the numbers $N_i$ correspond to the acceptance corrected meson yields.

\section{Results}
\label{sec:3}
\begin{figure}
\centering
\resizebox{0.45\textwidth}{!}{
  \includegraphics{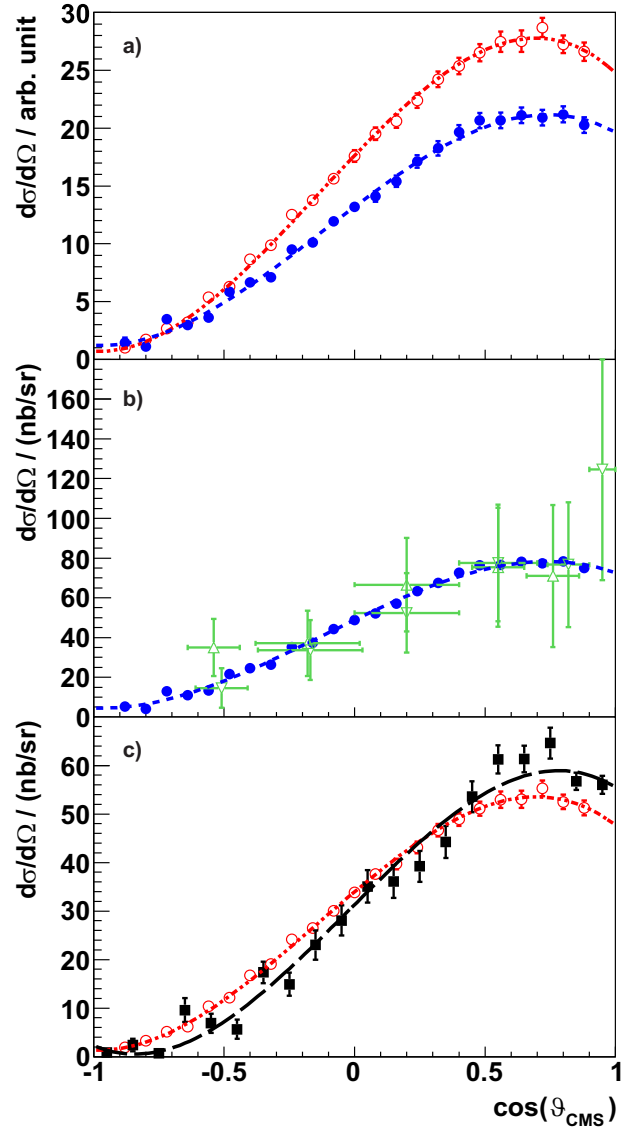}
}
\caption{(a) Angular distributions for the $48.8\,\mathrm{MeV}$ (filled
circles) and $59.8\,\mathrm{MeV}$ data (open circles) in arbitrary units.
(b) Angular distribution at $48.8\,\mathrm{MeV}$ from WASA-at-COSY
(filled circles) scaled to the data from GEM \cite{Betigeri2000}
(open triangles).
(c) Angular distribution at $59.8\,\mathrm{MeV}$ from WASA-at-COSY
(open circles) scaled to the $Q=59.4\,\mathrm{MeV}$ data from ANKE
\cite{Raus2009} (filled squares).
Curves correspond to third-order polynomial fits. The fit parameters are given
in table \ref{tab:parwasa}.}
\label{fig:diffcross}
\end{figure}

Figure \ref{fig:diffcross} (a) shows the angular distributions of the emitted
$\eta$ meson in the center-of-mass system in arbitrary units including statistical
uncertainties, systematic uncertainties due to the fitting of the missing-mass
spectra as well as the systematic uncertainties introduced by the absolute COSY
beam momentum precision of $0.1\,\%$ \cite{Maier1997}. The values for the
angular distributions are presented in table \ref{tab:diffcross}.

\begin{table*}
\centering
\caption{Extracted angular distributions for
$Q = (48.8 \pm 0.8)\,\mathrm{MeV}$ and $Q = (59.8 \pm 0.8)\,\mathrm{MeV}$ in
arbitrary units.}
\label{tab:diffcross}
\begin{tabular*}{0.88\textwidth}{@{\extracolsep{\fill}}p{1mm}cccp{1mm}}
\hline\noalign{\smallskip}
& $\cos{\vartheta_{\mathrm{CMS}}^\eta}$ &
\multicolumn{2}{c}{$\frac{\mathrm{d}\sigma}{\mathrm{d}\Omega} / \mathrm{arb.
unit}$} \\
&  & $Q = (48.8 \pm 0.8)\,\mathrm{MeV}$ & $Q = (59.8 \pm 0.8)\,\mathrm{MeV}$ \\
\hline\noalign{\smallskip}
& $-0.92$ - $-0.84$ & $1.43 \pm 0.47$ & $1.00 \pm 0.16$ \\
& $-0.84$ - $-0.76$ & $1.10 \pm 0.23$ & $1.71 \pm 0.20$ \\
& $-0.76$ - $-0.68$ & $3.49 \pm 0.22$ & $2.66 \pm 0.22$ \\
& $-0.68$ - $-0.60$ & $2.98 \pm 0.19$ & $3.21 \pm 0.12$ \\
& $-0.60$ - $-0.52$ & $3.61 \pm 0.22$ & $5.38 \pm 0.18$ \\
& $-0.52$ - $-0.44$ & $5.87 \pm 0.27$ & $6.29 \pm 0.20$ \\
& $-0.44$ - $-0.36$ & $6.66 \pm 0.29$ & $8.67 \pm 0.33$ \\
& $-0.36$ - $-0.28$ & $7.10 \pm 0.29$ & $9.89 \pm 0.34$ \\
& $-0.28$ - $-0.20$ & $9.51 \pm 0.35$ & $12.53 \pm 0.36$ \\
& $-0.20$ - $-0.12$ & $10.13 \pm 0.37$ & $13.77 \pm 0.39$ \\
& $-0.12$ - $-0.04$ & $11.95 \pm 0.41$ & $15.62 \pm 0.45$ \\
& $-0.04$ - $0.04$ & $13.21 \pm 0.44$ & $17.58 \pm 0.50$ \\
& $0.04$ - $0.12$ & $14.09 \pm 0.47$ & $19.52 \pm 0.55$ \\
& $0.12$ - $0.20$ & $15.40 \pm 0.50$ & $20.61 \pm 0.58$ \\
& $0.20$ - $0.28$ & $17.12 \pm 0.55$ & $22.38 \pm 0.63$ \\
& $0.28$ - $0.36$ & $18.26 \pm 0.62$ & $24.22 \pm 0.68$ \\
& $0.36$ - $0.44$ & $19.64 \pm 0.62$ & $25.38 \pm 0.71$ \\
& $0.44$ - $0.52$ & $20.66 \pm 0.66$ & $26.53 \pm 0.75$ \\
& $0.52$ - $0.60$ & $20.68 \pm 0.68$ & $27.49 \pm 0.88$ \\
& $0.60$ - $0.68$ & $21.12 \pm 0.67$ & $27.52 \pm 0.93$ \\
& $0.68$ - $0.76$ & $20.91 \pm 0.67$ & $28.71 \pm 0.84$ \\
& $0.76$ - $0.84$ & $21.19 \pm 0.69$ & $27.25 \pm 0.78$ \\
& $0.84$ - $0.92$ & $20.27 \pm 0.67$ & $26.62 \pm 0.78$ \\
\noalign{\smallskip}\hline
\end{tabular*}
\end{table*}

\begin{table*}
\centering
\caption{Parameters of third-order polynomial fits to angular distributions
determined at WASA-at-COSY and ANKE \cite{Raus2009}.}
\label{tab:parwasa}
\begin{tabular*}{0.88\textwidth}{@{\extracolsep{\fill}}p{1mm}llllllp{1mm}}
\hline\noalign{\smallskip}
& Experiment & $Q\ [\mathrm{MeV}]$ & $a_1$ & $a_2$ & $a_3$ & $\chi^2 / ndf$  \\
\noalign{\smallskip}\hline\noalign{\smallskip}
& WASA-at-COSY & $48.8 \pm 0.8$ & $1.300 \pm 0.028$ & $-0.211 \pm 0.029$ & $-0.60 \pm 0.06$ & $3.10$ \\
& WASA-at-COSY & $59.8 \pm 0.8$ & $1.337 \pm 0.017$ & $-0.277 \pm 0.024$ & $-0.65 \pm 0.04$ & $1.37$ \\ \hline\noalign{\smallskip}
& ANKE & $59.4 \pm 0.8$ & $1.72 \pm 0.06$ & $-0.08 \pm 0.05$ & $-0.87 \pm 0.08$ & $2.17$ \\
\noalign{\smallskip}\hline
\end{tabular*}
\end{table*}

The data can be described well by a third-order polynomial fit:
\begin{align}
\label{eq:pol}
\frac{\mathrm{d}\sigma}{\mathrm{d}\Omega}= a_0 \cdot \left[1+\sum_{n=1}^3 a_n \left(\cos{\vartheta_{\mathrm{CMS}}^{\eta}}\right)^n\right] \ .
\end{align}
The corresponding fit parameters $a_1$, $a_2$ and $a_3$ are given in table
\ref{tab:parwasa}.
Both data sets show a strong forward-peaked angular asymmetry, while the
backward cross sections almost vanish.

In the further analysis, extrapolations of the polynomial fits shown in
fig.~\ref{fig:diffcross} (a) are used to derive the yield for the two missing
angular bins in the very forward and backward direction. With these the cross
section ratio for the $\eta$ meson production has been determined to be:
\begin{align}
\label{eq:ratiocalc}
\frac{\sigma_{\eta}(48.8\,\mathrm{MeV})}{\sigma_{\eta}(59.8\,\mathrm{MeV})}
= 0.77 \pm 0.06 \ .
\end{align}
The uncertainty includes contributions from the number of
detected events as well as from acceptance determinations at $59.8\,\mathrm{MeV}$
excess energy ($0.5\,\%$) and at $48.8\,\mathrm{MeV}$ excess energy ($0.8\,\%$)
and uncertainties originating from the discussed normalization using the
$\mathrm{p}+\mathrm{d}\rightarrow {^3}{\mathrm{He}}+\pi^0$ reaction ($7.7\,\%$).

In fig.~\ref{fig:diffcross} (b) the new WASA-at-COSY $48.8\,\mathrm{MeV}$
angular distribution is compared to the GEM data \cite{Betigeri2000}
obtained at the same energy. Note that here the WASA-at-COSY data have been
scaled to the latter ones, \emph{i.e.} by a factor of
$3.70\,\mathrm{nb}/\mathrm{sr}$. Both data sets show the same angular asymmetry.
In fig.~\ref{fig:diffcross} (c) the $59.8\,\mathrm{MeV}$ WASA-at-COSY data are
compared to the $59.4\,\mathrm{MeV}$ ANKE data \cite{Raus2009}. The shown
WASA-at-COSY angular distribution is scaled to the ANKE data, \emph{i.e.} by a
factor of $1.93\,\mathrm{nb}/\mathrm{sr}$.
Also here both data sets show the same pronounced angular asymmetry.
However, due to the high integrated luminosity of the WASA-at-COSY experiment
the statistical uncertainties are significantly smaller than in the previous
experiments and allow for precise studies on the angular distributions.
\begin{figure*}
\centering
\resizebox{0.97\textwidth}{!}{
 \includegraphics{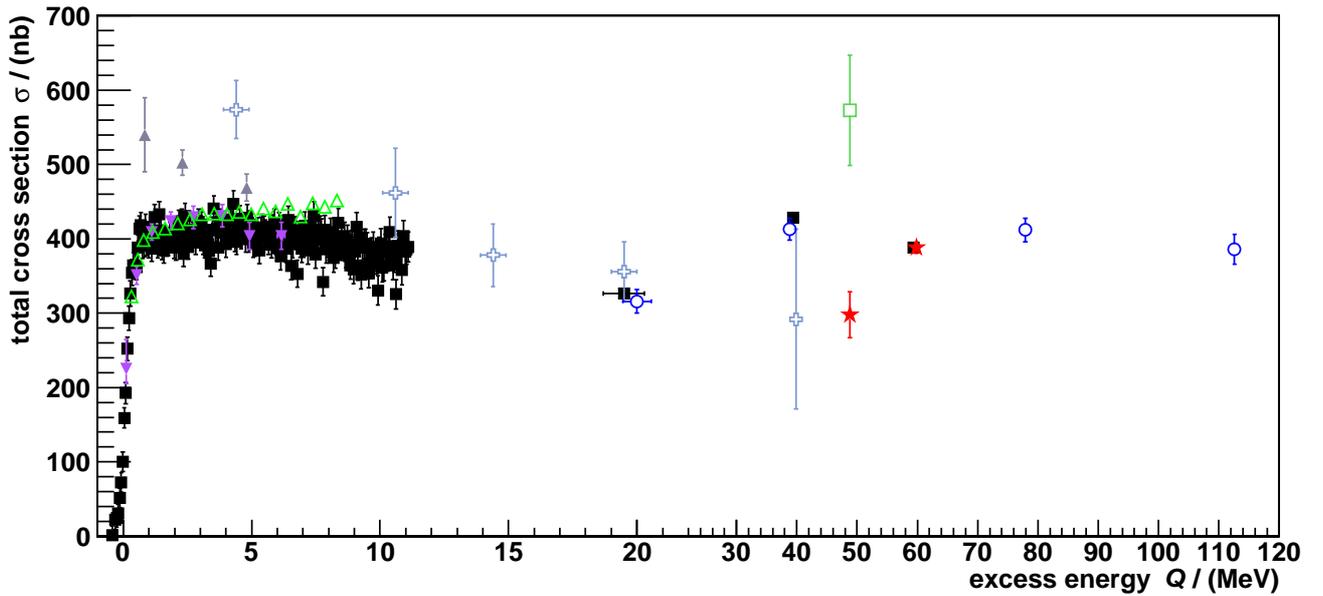}
}
\caption{Total cross sections of the reaction
$\mathrm{p}+\mathrm{d}\rightarrow {^3}{\mathrm{He}}+\eta$ as
function of the excess energy. All shown uncertainties are without systematic
uncertainties from absolute normalization. Besides our new result
(red stars) also shown are data from
ref.~\cite{Berger1988} (filled gray triangles),
ref.~\cite{Mayer1996} (inverted purple triangles),
ref.~\cite{Betigeri2000} (open green square),
refs.~\cite{Bilger2002,Bilger2004} (open blue circles),
ref.~\cite{Adam2007} (open light blue crosses),
ref.~\cite{Smyrski2007} (open green triangles)
and refs.~\cite{Mers2007,Raus2009} (filled black squares).
Here the data from WASA-at-COSY are arbitrarily scaled to the ANKE
data point at $Q = 59.4\,\mathrm{MeV}$. Note that the scale of the $Q$-axis
changes at $Q = 22\,\mathrm{MeV}$.} \label{fig:totcross}
\end{figure*}

In order to discuss the impact of the extracted cross section ratio the
$59.8\,\mathrm{MeV}$ data can be scaled to the cross section
$\sigma(59.4\,\mathrm{MeV}) = (388.1 \pm 7.1 \pm 58.0)\,\mathrm{nb}$ obtained at
ANKE \cite{Raus2009}. Here the latter uncertainty corresponds to the error in
the quoted absolute normalization of 15\%. In order to estimate systematic
uncertainties introduced by minor differences in the shape of the angular
distributions, both the ANKE and WASA data have been divided into five angular
bins to extract separate normalization factors for these intervals. The
uncertainty of the weighted mean value of these factors was determined to
$6.7\,\%$ and is used as systematic uncertainty. Using the scaling to the
$59.4\,\mathrm{MeV}$ data from ANKE and the ratio from eq.~(\ref{eq:ratiocalc})
one would derive a total cross section of
\begin{align}
\label{eq:totalcross}
\sigma (48.8 \pm 0.8\,\mathrm{MeV})= (298 \pm 24 \pm 49)\,\mathrm{nb} \ .
\end{align}
Here the first (statistical) uncertainty is dominated by the 
one of the determined ratio for the $\eta$ meson production ($7.7\,\%$).
In addition there are statistical uncertainties from the ANKE data point,
\emph{i.e.} $1.9\,\%$. The systematic uncertainty is dominated by the ANKE
data point ($15\,\%$) and the scaling to this data point ($6.7\,\%$).

In fig.~\ref{fig:totcross} this total cross section value is compared to the
existing data for the reaction
$\mathrm{p}+\mathrm{d}\rightarrow {^3}{\mathrm{He}}+\eta$ up to
$Q = 120\,\mathrm{MeV}$. Here it must be emphasized that our data are arbitrarily normalized relative to the ANKE data point at
$Q = 59.4\,\mathrm{MeV}$. As consequence the
cross section of the WASA-at-COSY data point at $Q = 59.8\,\mathrm{MeV}$
coincides with the one from ANKE. Note that all data points are presented
without absolute normalization uncertainties. Reason for this is that in the
excess energy region of interest, \emph{i.e.} above $Q = 20\,\mathrm{MeV}$, the
normalization uncertainties of the individual data sets are correlated.
Furthermore, the data from ANKE \cite{Raus2009} and WASA/PROMICE
\cite{Bilger2002,Bilger2004} which use the same reaction for the absolute
normalization are in very good agreement.
A similar argumentation holds for the close to threshold data where most of the
data originate from one single measurement from ANKE \cite{Mers2007} which
are normalized in the same way as the ANKE data at higher excess energies.
Obviously the precisely determined cross section ratio in combination with the
existing data base from ANKE \cite{Raus2009} and WASA/PROMICE
\cite{Bilger2002,Bilger2004} indicates the presence of a distinct
cross section variation between $Q = 20\,\mathrm{MeV}$ and
$Q = 60\,\mathrm{MeV}$, which is not smooth.

\section{Summary}
It turned out that the shape of the angular distributions obtained in the present
work agree well with those from previous measurements performed by the ANKE
collaboration \cite{Raus2009} and the GEM collaboration \cite{Betigeri2000}.
While at higher excess energies, \emph{i.e.} above $Q = 60\,\mathrm{MeV}$, the
excitation function exposes a smooth behavior, the new determined cross section
ratio $\sigma_{\eta}(48.8\,\mathrm{MeV}) / \sigma_{\eta}(59.8\,\mathrm{MeV})$
indicates the presence of a cross section variation in the region of
$Q = 20$--$60\,\mathrm{MeV}$.
Due to the comparatively high excess energy it is unlikely that this
effect is caused by a final state interaction. Instead, this observation
might be caused by the onset of higher partial waves or could indicate the
variation of the production mechanism. New total and differential cross section
data in the region of $Q = 20$--$80\,\mathrm{MeV}$ would be of high interest to
investigate this effect in more detail. An according measurement with the
WASA-at-COSY setup at COSY/J\"ulich was conducted in May 2014 \cite{{Khou2014}}.

\begin{acknowledgement}
This work was supported in part by the EU Integrated Infrastructure Initiative 
Hadron Physics Project under contract number RII3-CT-2004-506078; by the 
European Commission under the 7th Framework Programme through the ``Research 
Infrastructures'' action of the ``Capacities'' Programme, 
Call: FP7-INFRASTRUCTURES-2008-1, Grant Agreement Number 227431; by the Polish
National Science Centre through grant No. 2011/01/B/ST2/00431
and by the Foundation for Polish Science through the MPD programme.
We gratefully acknowledge the support given by the Swedish Research Council, 
the Knut and Alice Wallenberg Foundation, and the For\-schungs\-zen\-trum J\"ulich 
FFE Funding Program of the J\"ulich Center for Hadron Physics.
Finally we thank C.~Wilkin for the fruitful discussions.
\end{acknowledgement}

\end{document}